\documentclass[aps,twocolumn,prb,amsmath,amssymb,floatfix]{revtex4-2}
\usepackage{graphicx}
\makeatletter
\begin{document}
\title{Planckian dissipation and $c$-axis superfluid density in cuprate superconductors}

\author{Sudip Chakravarty}
\affiliation{Mani L. Bhaumik Institute for Theoretical Physics\\Department of Physics and Astronomy, University of
California Los Angeles, Los Angeles, California 90095-1547}

\pacs{}
\begin{abstract}
An interesting concept in condensed matter physics is Planckian dissipation, in particular its manifestation in a remarkable phenomenology of superfluid density as a function of superconducting transition temperature.  The concept was ontroduced for $ab$-plane properties.  However, when suitably interpreted,  it can  also be applicable to the incoherent $c$-axis resistivity, which has not been adequately  addressed previously. There are two results in this note: the first  is a derivation  using Kubo formula as to how Planckian dissipation could arise. It is aided by the fact that the $c$-axis tunneling matrix element is so small that a second order perturbation theory combined with presumed non-Fermi liquid behavior is sufficient to illuminate the phenomonon. In addition, the notion of quantum criticality  plays an important role. 
\end{abstract}
\maketitle
 \section{Introduction}
 
Many mysteries of high temperature cuprate superconductors are still unresolved after decades of their discoveries. In this paper one such mystery and its related concepts are considered in the hope that it might  inspire continued thought. Here I would like to consider  Planckian dissipation, so coined by Jan Zaanen~\cite{Zaanen:2004}. This idea has its roots in the observation of a correlation between the $ab$-plane  resistivity  and the corresponding superfluid density  for remarkably large range of transition temperatures of the cuprate superconductors. However, that the  $c$-axis  resistivity (perpendicular to the $ab$-plane) also shows scaling with the $c$-axis superfluid density  has not been fully addressed. I am going to focus on  this scaling.

The idea of Zaanen is a dimensional argument in which the dimensions of  the observed correlation between the $ab$-plane superfluid density, $\rho_s$,  and the corresponding $ab$-plane conductivity,  $\rho_s =  k_BT_c \sigma_{ab} (T_c)$, has to be correctly 
matched~\cite{Homes:2004}. While the left hand side has the dimension $1/s^2$ ($s\equiv \text{second}$), the right hand side will carry the same dimension if energy is   also expressed as $1/s$ and this requires Planck's constant $\hbar$; $\sigma_{ab}$ already has the dimension of $1/s$. The correlation  therefore is  a truly  quantum phenomenon. There is little to contest because dimensional analysis is of absolute validity; the timescale $\tau_c= \hbar/k_B T_c$, Planckian time, is very short. Any shorter, it will be dissipationless superfluid at all temperatures!  This may be one reason why the temperature is so high, as remarked by Zaanen~\cite{Zaanen:2004}. Although this argument reveals a dissipative time scale, it does not provide any reason why such a relation will theoretically hold. It is a description of the experimentally observed data,~\cite{Legros:2019} notwithstanding the later paper by Zaanen~\cite{Zaanen:2019}.
Planckian dissipation, as reflected in the linear resistivity in  correlated materials, is extensively reviewed by Hartnoll and Mackenzie~\cite{Hartnoll:2022}. The discussion, however,  is for the $ab$-plane resistivity in the layered cuprate superconductors, not for the $c$-axis resistivity. 

Attempts, however,  have been made in the experimental literature~\cite{Homes:2004}  to correlate the   $c$-axis superfluid density to the  $c$-axis resistivity,  but it remains unconvincing because of the use of the Fermi liquid language, which  may not be applicable.   However, I will summarize briefly the chain of reasoning~\cite{Homes:2004}  for the sake of completeness. In the Ambegaokar-Baratoff (AB) formula~\cite{Ambegaokar:1963}  we set $T=0$ and define $c$-axis superfluid density in terms of the  the gap proportional to  $T_c$.  Then by considering the tunneling resistance $R_N = \rho_c d$  one obtains a result similar to the $ab$-plane result discussed above; here $d$ is the spacing between the layers and $\rho_c$ is the $c$-axis resistivity.  

From what we know now about electron correlations in high temperature superconductors, this argument is  deficient, and the validity of AB formula can be  questioned, as was pointed out  by Chakravarty and Anderson~\cite{Chakravarty:1994}.   In addition, the $c$-axis resistivity has quite often  a semiconducting upturn before $T_c$ is reached.~\cite{Boebinger:1996} Another complicating factor is that the gaps in the cuprates are of $d$-wave character, not $s$-wave, and therefore additional assumptions must be made.

The purpose of this paper is to examine  Planckian dissipation in the context of  quantum criticality~\cite{Chakravarty:1988,Chakravarty:1989} and incoherent $c$-axis tunneling in a strongly correlated electron system.

\section{c-axis conductivity sum rule}
Consider a model in which the microscopic
Hamiltonian  of electrons hopping along the $c$-axis is
\begin{equation}
H_c=-t_{\perp}\sum_{jl,s}c^{\dagger}_{jl,s}c_{jl+1,s}+{\rm h. c.},
\end{equation}
where the label $j$ refers to the sites of the two-dimensional plane, $l$ refers to the layer
index, and $s$ refers to spin; $c^{\dagger}_{jl,s}$ is the electron creation operator. The hopping along $c$-axis could be of extended nature without losing the generality of our argument.
The complete Hamiltonian is therefore
\begin{equation}
H=H_{\rm rest}+H_c
\end{equation}
$H_{\rm rest}$ is arbitrary, but does not contain any interplane hopping. The sum rule derived below remains unchanged if interplane interactions
that do not transfer electrons between the planes
are allowed, as mentioned above.
Here I
focus  only on  single layer materials for which all CuO-planes
are equivalent, such as in LSCO, Tl2201, Hg1201, Bi2201, etc.. 
The frequency and 
wavevector dependent $c$-axis conductivity can be written as~\cite{Shastry:1990}
\begin{equation}
\sigma^c(q_c,\omega,T)=-{1\over Ad}\left({ed\over \hbar}\right)^2 \frac{\langle
-H_c(T)\rangle-\Lambda^c_{\rm ret}(q_c,\omega,T)}{i(\omega+i\delta)},
\end{equation}
where $q_c$ is the momentum transfer perpendicular to the plane, $A$ is the two-dimensional area and $d$
is the separation between the layers. The retarded current-current commutator is 
\begin{equation}
\Lambda^c_{\rm ret}(l,t,T)=-i\theta(t) \langle [j_{H}^c(l,t),j_{H}^c(0,0)]\rangle. 
\end{equation}
The paramagnetic current operator is defined by
\begin{equation}
j^c(l)=it_{\perp}\sum_{j}(c^{\dagger}_{jl,s}c_{jl+1,s}- {\rm h. c.}),
\end{equation}
and the corresponding Heisenberg operator, $j^c_{H}$, is defined with respect to the full 
 Hamiltonian. The
averages refer to the thermal averages and 
$
\langle H_c(T)\rangle = -t_{\perp}\sum_{j,s}\langle c^{\dagger}_{jl,s}c_{jl+1,s}+{\rm h. c.}\rangle.
$
For optical conductivity, one may set $q_c=0$, and, then noting that the retarded
current-current commutator  is analytic in the upper-half of the complex $\omega$-plane, we arrive at an {\em exact}
$c$-axis conductivity sum rule, originally derived by Kubo~\cite{Kubo:1957}
\begin{equation}
\int_{-\infty}^{\infty}d\omega\ {\rm Re}\ \sigma^c(\omega,T)={\pi e^2 d^2\over \hbar^2 Ad}\langle
-H_c(T)\rangle,
\label{srule}
\end{equation}
which is a variant of  the well-known $f$-sum rule.~\cite{Tinkham:1996}  

The right hand side of the sum rule is the average of the tunneling 
Hamiltonian. This may be deceptively simple because it is the true $c$-axis interacting kinetic energy. 
 The absence of Galilean invariance on a lattice  allows the charge carrying effective mass to vary with
temperature and interaction. In the continuum limit, such that $d\to 0$, but  $t_{\perp}d^2$ fixed, the
right hand side of Eq.~(\ref{srule}) is ${\pi n e^2\over m}$, where
${\hbar^2\over 2m} = t_{\perp}d^2$, and $n$ is the density of electrons in the planes. In this limit, interactions cannot renormalize the effective mass because the
current operator commutes with the Hamiltonian. 
  
Because  
$H_c$ is a  low-energy effective hamiltonian, the upper 
limit of the integral in Eq.~(\ref{srule})  cannot exceed 
an inter-band cutoff $\omega_c$, of the order of a few electron volts. Beyond this we need not be more specific, because our goal is to 
deduce results as $T\to 0$. 

We now evaluate the sum rule in what follows. For a superconductor, we can write quite generally
\begin{equation}
{\rm Re}\ \sigma^{cs}(\omega,T)=D_c(T)\delta(\omega)+{\rm Re}\ \sigma^{cs}_{\rm reg}(\omega,T).
\end{equation}
The first term signifies the lossless  flow of electrons in the superconducting state, while the second
is the regular (nonsingular) part of the optical conductivity. 
The normal state optical conductivity is nonsingular; so,
the sum rule can be cast into a more useful form:
\begin{eqnarray}
D_c(T)&=&\int_0^{\infty}d\omega \bigg[{\rm Re}\ \sigma^{cn}(\omega,T)-{\rm Re}\ \sigma^{cs}_{\rm
reg}(\omega,T)\bigg]\nonumber\\
&+&{\pi e^2 d^2\over 2 Ad \hbar^2}\bigg[\langle -H_c(T)\rangle_s-\langle -H_c(T)\rangle_n\bigg]
\label{srule2}
\end{eqnarray}

If the $c$-axis kinetic energy is unchanged between the normal and the superconducting
states, as  in a conventional layered superconductor, we recover a variant of the
Ferrell-Glover-Tinkham sum rule.~\cite{Tinkham:1996}
The missing area between the $c$-axis conductivities of the normal and the superconducting states 
is proportional to the $c$-axis superfluid density.

The $c$-axis  resistivity is generically semiconducting and strongly temperature dependent, at least in the
underdoped and optimally doped regimes. This temperature dependence should persist
if superconductivity could be suppressed, say by applying a magnetic field, and therefore 
the  equality of  the conductivity at $T_c$ and those at  $T\le T_c$ cannot be assumed. For LSCO, this has been
demonstrated experimentally~\cite{Ando:1995}.  

On general grounds, there is  little 
we can say about ${\rm Re}\ \sigma^{cn}(\omega,T)$ for $T\le T_c$. 
To proceed further consider the sum rule at zero temperature, 
which can be restated as 
\begin{equation}
D_c(0)\ge {\pi e^2 d^2\over 2 Ad\hbar^2}\left[\langle -H_c(0)\rangle_s-\langle -H_c(0)\rangle_n\right].
\end{equation}
I have assumed that the integral in Eq.~(\ref{srule2}) is positive definite.
This could
be a strict inequality, although I cannot find a rigorous  argument. One can see, however, that at very high
frequencies the two conductivites should approach each other, and, at low frequencies, $\sigma^{cn}_{\rm
reg}(\omega,T=0)\ge \sigma^{cs}_{\rm reg}(\omega,T=0)$, if the superconducting state is at least
partially gapped. 

If the experiments of Ando {\em et al.}\cite{Ando:1995} are taken as a serious  indication, the system, at $T=0$, is
insulating along the $c$-axis. It is plausible, therefore, that  the integral in Eq.~(\ref{srule2}) is
smaller than what one would have guessed for metallic conduction along the $c$-axis. This is because the
frequency dependent $c$-axis conductivity in a non-Fermi liquid is expected to vanish as a power law in
contrast to the Drude behavior.
If this is indeed true, we can make the approximation
\begin{equation}
D_c(0)\approx{\pi e^2 d^2\over 2 Ad\hbar^2}\left[\langle -H_c(0)\rangle_s-\langle
-H_c(0)\rangle_n\right].
\label{Dc0}
\end{equation}
The $c$-axis penetration depth is given by
\begin{equation}
{1\over\lambda_c^2(0)}={8D_c(0)\over c^2},
\end{equation}
where $c$ is the velocity of light.

\section{Hopping along $c$-axis}
To estimate the right hand side of Eq.~(\ref{srule}), one must make a distinction between Fermi and non-Fermi liquids. While there 
is no universal theory of non-Fermi liquids, its characterization in terms of spectral function is unambiguous.We would like to build a model in which the non-Fermi liquid behavior is 
 essential, which we briefly recall: the  analytic
continuation of the Green's function to the second Riemann sheet should contain branch points instead
of simple poles.~\cite{Yin:1996}A spectral function $A$  that satisfies the
scaling relation
\begin{equation}
A(\Lambda^{y_1}[k-k_F],\Lambda^{y_2}\omega)= \Lambda^{y_A}A([k-k_F],\omega),
\end{equation}
where $y_1$, $y_2$, and $y_A$ are the  exponents defining the universality class of
the critical Fermi system. The values of the exponents other than the set
$y_1=1$, $y_2=1$, and $y_A=-1$ 
represent a non-Fermi liquid by our definition.  

For a  non-Fermi liquid, an electron creation operator of wavevector {\bf k} and spin $\uparrow$ acting
on the ground state  creates a linear {\em superposition of states} that carry the momentum $\bf k$ and spin $\uparrow$. 
However, the act of
inserting an electron into a non-Fermi liquid cannot be renormalized away by defining a single
quasiparticle:    the excitation energy is not uniquely related to a given $\bf k$. 

The average $\langle H_c \rangle$ is to lowest order is $t_{\perp}^2/W$ 
in a perturbative expansion.
To prove that note that the {\em exact} ground state  expectation value in terms of $\widehat 0$ is given by 
\begin{equation} 
\langle \widehat{0}|H_c|\widehat{0}\rangle=-2\sum_{n\ne 0}\frac{|\langle 
0|H_c|n\rangle|^2}{E_n-E_{0}}+O(t_{\perp}^4), 
\label{secondorder}
\end{equation} 
where $E_n$ and $|n\rangle$ on the right are the eigenvalues and the eigenstates  
of the Hamiltonian without the $c$-axis tunneling Hamiltonian. The first order term is zero, because of the absence  
of coherent single particle tunneling between the unit cells,  which is a non-fermi liquid feature of the in-plane excitations.
Thus,  the energy 
denominator can be approximated by $W$, and the sum can 
be collapsed using the completeness 
condition to 
$\langle 0|H_c^2|0\rangle/W$. Thus the effective hamiltonian is $-H_c^2/W$. 

There is an alternative way to motivate the same conclusion. Of course, neither of them are rigorous.
We perform a canonical 
transformation such that 
$H_c$ is eliminated from the hamiltonian $H=H_{\rm rest}+H_c$. Thus, 
\begin{equation} 
\tilde{H}=e^{-S}He^{S}=H_{\rm rest}+{1\over 
2}[H_c,S]+\cdots,\label{Htilde} 
\end{equation} 
where the antihermitian operator $S$ is defined by $H_c+[H_{\rm 
rest},S]=0$. 
The ground state  $|\tilde{0}\rangle$ of the full hamiltonian $H$ can be 
determined perturbatively in $S$ (or, equivalently $t_{\perp}$) to show 
that the result is the same as above.

For conserved parallel momentum, the expansion on the right hand side of 
Eq.~(\ref{secondorder}) does not converge in a Fermi liquid  theory because of 
vanishing energy denominators; therefore the expansion would not be 
valid. In  a gapped  state,  
the expansion can be  legitimate because of the absence of vanishing energy 
denominators. 
In a non-Fermi liquid state, the matrix 
elements should vanish  for 
vanishing energy differences, and  the the sum is skewed to 
high energies. Thus,  the energy 
denominator can be approximated by $W$, and the sum can 
be collapsed using the completeness 
condition to $\langle \tilde{0}|H_c|\tilde{0}\rangle\approx -\langle 
0|H_c^2|0\rangle/W$. Thus $\langle H_c \rangle$ is of order $t_{\perp}^2/W$.

\section{Analysis of the sum rule}

Returning to Eq.~\ref{srule} one finds on dimensional grounds that the $c$-axis conductivity can be expressed as  
\begin{equation} 
\sigma_c(T)=a\left(\frac{e^2 d t_{\perp}^2}{ A W \hbar^2}\right){\frac{1}
{\Omega(T)}}, 
\label{conductivity} 
\end{equation} 
where $a$ is a numerical constant weakly dependent on the band 
structure. Thus, the inelastic 
scattering rate is proportional to an {\em unknown} function $\Omega(T)$. 
From a simple manipulation,  the $T=0$
$c$-axis penetration depth ($\lambda_{c}$) is given by, using Eq.~\ref{Dc0},
\begin{equation} 
\frac{c^2}{8 \lambda_c^2}=\frac{4\pi}{a}  \sigma_c(T) \Omega(T) 
\left[u_s-u_n\right], 
\label{penetration} 
\end{equation} 
where $u_{s,n}$, clearly dimensionless quantities, are  
\begin{equation}
u_{s,n} = \langle0|(H_c/t_{\perp})^2|0\rangle_{s,n}
\end{equation}
Note that the  average here is 
with respect to the ground state of the Hamiltonians without the $c$-axis piece, that is only  the planar parts. The subscripts $s$ and $n$ refer to  superconducting and  normal ground states. By normal  state I mean a state in which superconductivity is destroyed. Presently it can be achieved reasonably well by applying high magnetic field~\cite{Taillefer:2007}. Only temperature dependencies  are contained in the product $\sigma_c(T) \Omega(T)$.

\begin{figure}[hbtp]
\includegraphics[width=\linewidth]{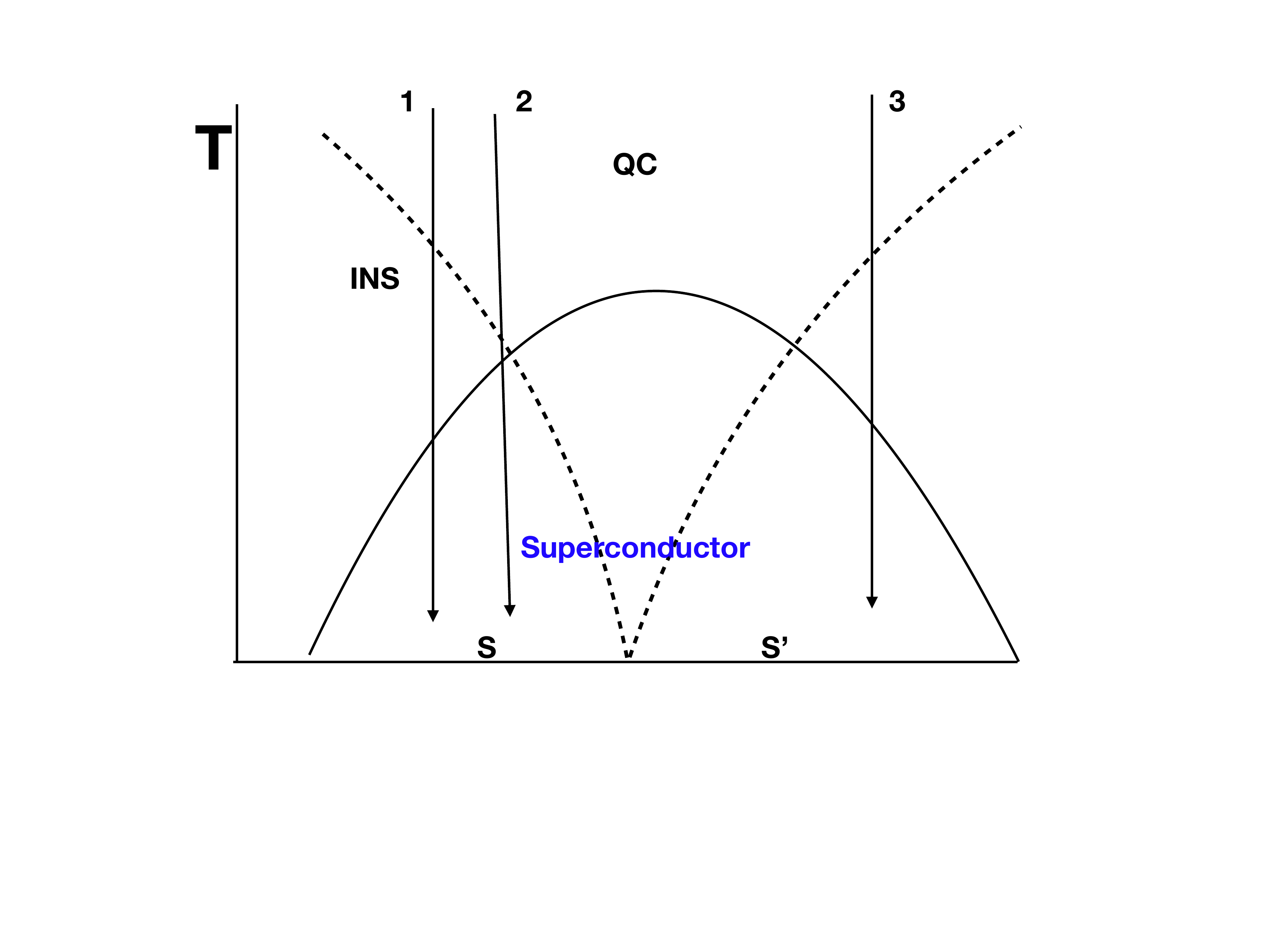}
\caption{A rough sketch of dimensional crossover of $\rho_{c}$ as a function of temperature $T$. the dashed lines schematically represents quantum critical fans ant the solid dome is the superconducting dome. }
\label{phase}
\end{figure}

To make progress, we note that the $c$-axis resistivity, 
$\rho_c(T)$, can often be fitted   to a combination of power laws ($p>0$). This power law behavior is not necessary as long as there is a quantum critical crossover, as we shall discuss below. The present form is chosen for convenience, let
\begin{equation} 
\rho_c(T)=b_1 T^{-p}+b_2'T. 
\label{caxisrho}
\end{equation} 
In some materials such as $\mathrm{YBa_2Cu_{3}O_{6+\delta}}$ the $c$-axis resistivity may not be substantially larger 
from the $ab$-plane resistivity close to $T_{c}$ and may not have an insulating  upturn. From the picture, see line 2 in Fig.~\ref{phase}, it is easy to understand  to be the case when  the the experimental trajectory is very close to the quantum critical fan and the superconducting boundary. 

We now rewrite Eq.~(\ref{conductivity}) as,
\begin{equation} 
\Omega(T)=a\left(\frac{e^2 d t_{\perp}^2}{A W \hbar^2}\right) \rho_c(T)
\end{equation}
This equation allows us to determine the unknown $\Omega(T)$.
We express 
Eq.~(\ref{penetration}) in terms of the temperature $T^*$ (not the notation for the pseudogap) at which the 
c-axis resistivity takes its minimum value given the empirical behavior, Eq.~\ref{caxisrho} .
We get 
\begin{equation} 
\frac{c^2}{\lambda_c^2}=\sigma_c(T^*) T^* 
\left\{4\pi \frac{b_2(p+1)}{ p}\left[u_s-u_n\right]\right\}, 
\label{Eq8}
\end{equation} 
where $b_2=b_2'(d e^2 t_{\perp}^2/\hbar^2 A W)$. 
The expression in the curly brackets depends dominantly on $b_2$, 
which describes the high temperature linear
resistivity. The low temperature behavior enters only through the 
exponent 
$p$, but of course cuts off at $T_c$. 

What could be the meaning of $T^*$? At a trivial level it is a
lower bound to $\rho_c$. It is also the boundary of the quantum critical region above which the linear behavior 
of the resistivity appears.
INS stands for insulator and QC for quantum critical. Here S and S' are two phases, both with superconducting order parameters. For example, S may be $d$-density wave along with superconductivity and S' without the $d$-density wave.  Thus there can be a continuous phase transition. Three vertical arrows are possible experimental trajectories that intersect with quantum critical fan describing a crossover scale. These trajectories cross over to the superconducting state. {\em There is no reason for the quantum critical point to be in the middle of the superconducting dome as drawn here for simplicity.} The line 2 will have  very little insulating up turn.

\begin{figure}[htbp]
\includegraphics[width=\linewidth]{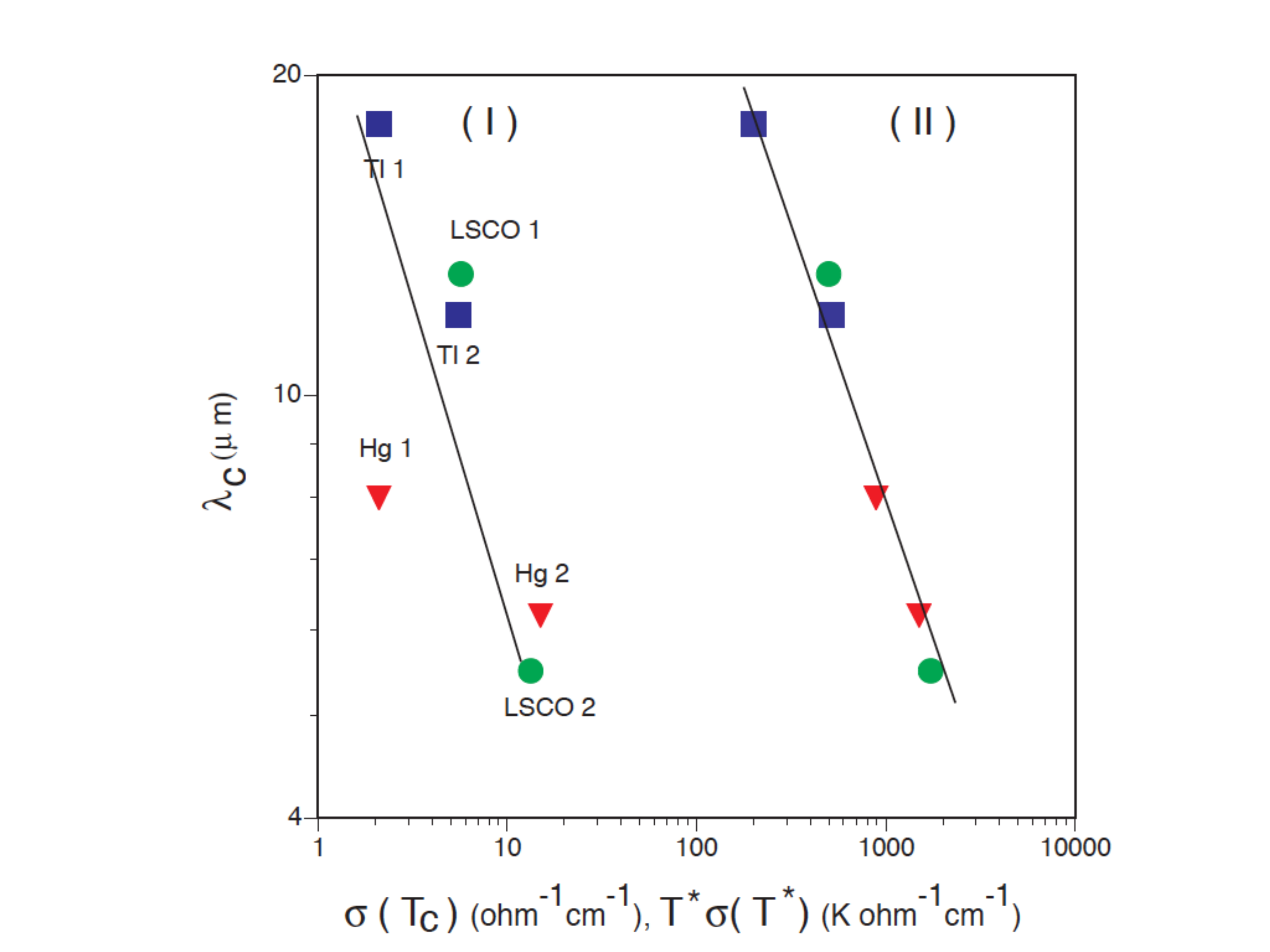}
\caption{The Basov plot: $\ln \lambda_c$ is plotted against  $\ln 
\sigma_c(T_c)$, as in the original 
Basov plot (I), and against $\ln [T^*\sigma_c(T^*)]$ as discussed here 
(II). The legends in group (II) are the same as those in group (I). Tl1: Ref.~\cite{Moler:1998} Tl2: Ref.~\cite{Basov:1999}; 
Hg1: Ref.~\cite{Kirtley:1998}, Hg2: Ref.~\cite{Basov:1999}; LSCO 1 (12\%), LSCO 2 (15\%): Refs.~\cite{Loram:1994,
Uchida:1996,Nakamura:1993}. More recent compilation of data~\cite{Homes:2004} does not change any conclusions.}
\label{Fig1}
\end{figure}
The expression in Eq.~\ref{Eq8} is very interesting. It is cast in the form discussed by Zaanen~\cite{Zaanen:2004} and shows  the importance of 
Planckian time $\hbar/ k_BT^*$, once again by matching the dimensions.
There are two important points worth noting. First $T^*$ is not the same as $T_c$, the transition temperature; it is merely a characteristic temperature. Second, provided the expression in curly brackets is a universal 
constant, a plot of $\ln \lambda_c$ against 
$\ln[\sigma_c(T^*) T^*]$ should be a universal straight line, 
independent of material, and temperature $T^{*}$, which is {\em indeed} the case, thus validating Eq.~(\ref{Eq8}).
Basov {\em 
et al.}\cite{Basov:1994} suggested a similar correlation by plotting 
$\ln \lambda_c$ against 
$\ln\sigma_c(T_c)$, shown as (I) in Fig.~\ref{Fig1}. In comparison to Basov correlation, the correlation 
discussed here, shown as (II), is excellent. 
In Fig.~\ref{Fig1}, we have taken $T^*\approx T_c$ 
for those  doped 
materials that show simply a flattening of $\rho_{c}(T)$ close to $T_c$; see the discussion above in reference to Fig.~\ref{phase}.  

\section{Discussion}Recall that in order to discuss Planckian dissipation above we had to introduce a sum rule due to  Kubo. This sum rule applies independently for any direction. The  $c$-axis sum rule has not been discussed in this context to my knowledge. It is notable that I could combine a simple perturbative expansion, thanks to the  assumption of lack of quasiparticle pole,  with some ``empirical'' results concerning the $c$-axis resistivity, which is physically clearly justified  in Fig.~\ref{phase}. I could have used the $ab$-plane sum rule but in that case a perturbative analysis would not have been applicable because of strong interactions.  Later I relied on quantum criticality to arrive at the dissipation scale. Is there any interpretation that we can give to  the $T=0$ problem above?  A rigorous answer does not exist.   I finish by emphasizing that the existence of quantum criticality does not always lead to time scales relevant for  dissipation, as discussed in regard to compactification scale~\cite{Chakravarty:1996}.  Perhaps one can extend much of the theory to the criticality in the overdoped regime of cuprates~\cite{Kopp:2007,Legros:2019}, recently discussed in Refs.~ \cite{Kurashima:2018} and~\cite{Sarkar:2020}.

A number of issues are not addressed: I only considered single layer cuprates and said nothing about the multilayer cuprates. Moreover the treatment in the $c$-axis tunneling in a non-Fermi liquid has been heuristic. These difficult questions must be answered in the future.

This work was performed at the Aspen Center for Physics, which is supported by National Science Foundation grant PHY-1607611.
It  was also partially supported by funds from David S. Saxon Presidential Term Chair at the University of California Los Angeles. I thank Steve Kivelson for a crtical reading of the manuscript.

%

\end{document}